\documentstyle[aps,twocolumn,eqsecnum,psfig]{revtex}

\title{Coupling Impedances of Azimuthally Symmetric Obstacles \\
of Semi-Elliptical Shape in a Beam Pipe\thanks{Work supported by 
   the U.S.\ Department of Energy}}
\author{Robert~L.~Gluckstern and Sergey~S.~Kurennoy\thanks{Present 
   address: AOT-1, LANL, Los Alamos, NM 87545} \\
 {\it Physics Department, University of Maryland, 
      College Park, MD 20742}}

\begin{document}
\maketitle

\begin{abstract} 
The beam coupling impedances of small axisymmetric obstacles having 
a semi-elliptical cross section along the beam in the vacuum chamber 
of an accelerator are calculated at frequencies for which the 
wavelength is large compared to a typical size of the obstacle. 
Analytical results are obtained for both the irises and the cavities 
with such a shape which allow simple estimates of their broad-band 
impedances. 
\end{abstract} 

\narrowtext

\section{Introduction} 
High currents in modern accelerators and colliders severely restrict 
the allowed coupling impedance of the machine. For this reason, it 
is important to know the impedance contributions even from small 
discontinuities of the vacuum chamber.

In a recent paper \cite{1}, Kurennoy has analytically calculated the
low-frequency coupling impedance of small obstacles protruding into 
a beam pipe. In this paper we present an alternative derivation for 
an azimuthally symmetric semi-elliptical object protruding into a 
beam pipe, which confirms the dependence on the depth, but not on 
the width, of the protrusion. We also study the more difficult --- 
from the analytical point of view --- case of an axisymmetric 
semi-elliptical protrusion {\em outside} the beam pipe (cavity),
and present variational results for different elliptical eccentricities.

\section{General Analysis}
Consider a beam pipe of radius $R_{\rm pipe}$ and an azimuthally 
symmetric obstacle whose dimensions are small compared with both 
$R_{\rm pipe}$ and $\lambda$, the rf wavelength.  We start with the 
definition of the longitudinal impedance as \cite{2}
\begin{equation}
Z_{\parallel} (k) = \frac{1}{|I_0|^2} \int dv \vec{E} \cdot J^*,
\end{equation}
where the current in the frequency domain for an ultrarelativistic 
point charge is
\begin{equation}
J_z (x,y,z;k) = I_0 \delta (x) \delta (y) \exp (-jkz),
\end{equation}
with $k = w/c = 2\pi /\delta$, and with the implied time dependence 
of all quantities being $\exp (jwt)$.  We then identify two 
configurations: \ the subscript 1 denotes the pipe without the 
obstacle and the subscript 2 denotes
the pipe with the obstacle.  By forming the combination
\[ 
-\int dv (\mbox{\boldmath $E$}_2 \cdot \mbox{\boldmath $J$}^* +
\mbox{\boldmath
$E$}_1^* \times \mbox{\boldmath $J$})
\]
and using Maxwell's equations to write $\mbox{\boldmath $J$}$ in 
terms of the fields $\mbox{\boldmath $E$}$ and $\mbox{\boldmath $H$}$, 
we write the contribution of the obstacle to the impedance as
\begin{equation}
|I_0|^2 Z_u(k) = \int_{S_2 \neq S_1} dS_2 \mbox{\boldmath $n$}_2 \cdot
\mbox{\boldmath $E$}_1^* \times \mbox{\boldmath $H$}_2,
\end{equation}
where the surface integral is only over the surface of the obstacle.  
Using
\begin{equation}
E^*_{1r} = \frac{Z_0I_0^*}{2\pi r} \exp (jkz)
\end{equation}
and 
\begin{equation}
\mbox{\boldmath $n$}_2 dS_2 = 2\pi r [\hat{\mbox{\boldmath $n$}} dz -
\hat{\mbox{\boldmath $z$}} dr],
\end{equation} 
with $\hat{\mbox{\boldmath $r$}}$ and $\hat{\mbox{\boldmath $z$}}$ 
being unit vectors, we have
\begin{equation}
\frac{Z_{\parallel}(k)}{Z_0} = -\frac{1}{I_0} \int dr H_{2\phi} e^{jkz}.
\end{equation}

Figure~1 show the geometry for an obstacle protruding into and outside 
of the beam pipe.

\begin{figure}[htb]
\centerline{\psfig{file=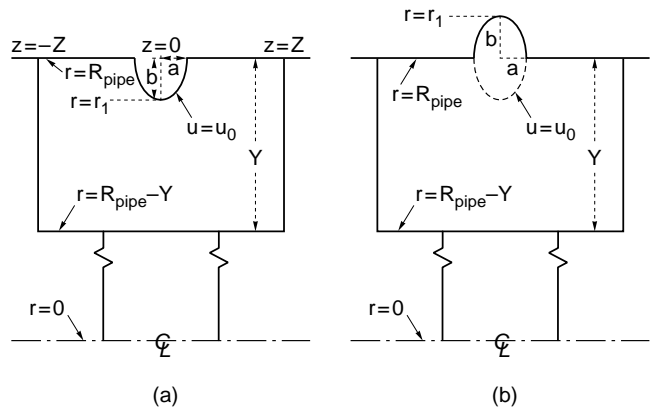,width=8.5cm}}
\caption{The beam pipe with an interior (a) and exterior (b) 
obstacle. One assumes $a,b \ll Y,Z \ll \lambda, R_{\rm pipe}$.}
\end{figure}

A more explicit form for Eq.~(2.6) is
\begin{eqnarray}
\frac{Z_{\parallel}(k)}{Z_0} &=& -\frac{1}{I_0} 
\left[ \int^{r_1}_{R_{\rm pipe}}
dr H_{2\phi} (r,z_a(r)) e^{jkz(r)}\right. \nonumber \\
&+& \left. \int^{R_{\rm pipe}}_{r_1} dr H_{2\phi} (r,z_b(r)) 
e^{jkz(r)}\right]
\end{eqnarray}

We now convert the bracket in Eq.~(2.7) to a double integral for the
obstacle in Fig.~1a:
\begin{eqnarray}
[ \ ] &=& \int^{R_{\rm pipe}}_{r_1} \int^{z_a(r)}_{-Z} dz
\frac{\partial}{\partial z} (H_{2\phi} e^{jkz}) \nonumber \\
&+& \left. \int^{R_{\rm pipe}}_{r_1} dr (H_{2\phi} e^{jkz}) \right|_{z=-Z}
\nonumber \\  
&+& \int^{R_{\rm pipe}}_{r_1} dr \int^{Z}_{z_b(r)} dz
\frac{\partial}{\partial z} (H_{2\phi} e^{jzk}) \nonumber \\
&-& \left. \int^{R_{\rm pipe}}_{r_1} dr (H_{2\phi} e^{jkz}) \right|_{z=Z}. 
\end{eqnarray}
Here $Z$ is a distance large compared with the dimensions of the 
obstacle, but small compared with $\lambda$ and $R_{\rm pipe}$, so 
that $H_{2\phi} \exp (jkz)$ takes on its value in a pipe without an 
obstacle at $z = \pm Z$, namely
\begin{equation}
H_{2\phi} (\pm Z) = I_0 \exp (\mp jkZ).
\end{equation}
This causes the 2nd and 4th terms on the right side of Eq.~(2.8) to 
cancel. We also add the vanishing term
\begin{equation}
\int^{r_1}_{R_{\rm pipe}-Y} dr \int^Z_{-Z} dz \frac{\partial}
{\partial z} (H_{2\phi} e^{jkz}),
\end{equation}
and finally obtain
\begin{equation}
\frac{Z_{\parallel}(k)}{Z_0} = \frac{1}{I_0}
\int \! \! \int_{\stackrel{\mbox{\scriptsize solid}}
{\mbox{\scriptsize border}}} drdz
\frac{\partial}{\partial z} (H_{2\phi} e^{jkz}),
\end{equation}
where the area of integration is within the solid border in Fig.~1a. 
Parallel arguments lead to the same result for the obstacle in Fig.~1b.

We now use Maxwell's equations to rewrite Eq.~(2.11) as
\begin{equation}
\frac{Z_{\parallel}(k)}{Z_0} = -\frac{jk}{Z_0I_0} \int \! \!
\int_{\stackrel{\mbox{\scriptsize solid}}{\mbox{\scriptsize border}}} 
 drdz (E_r - Z_0H_{\phi}) e^{jkz}
\end{equation}
where we have dropped the subscript 2.  For low frequency we set $\exp
(jkz)=1$ and obtain
\begin{equation}
\frac{Z_{\parallel}(k)}{Z_0} \cong -\frac{jk}{Z_0I_0} \int \! \!
\int_{\stackrel{\mbox{\scriptsize solid}}{\mbox{\scriptsize border}}} 
drdz (E_r - Z_0H_{\phi}).
\end{equation}
Clearly our derivation has resulted in a separation into a term 
involving the electric polarizability and a term involving the 
magnetic susceptibility (in the azimuthal direction) as in previous 
work \cite{3,4,5,6}.  Since the obstacles are azimuthally symmetric, 
we can replace $Z_0H_{\phi}$ in the vicinity of the obstacle by
\begin{equation}
Z_0H_{\phi} = E_0 = Z_0I_0/(2\pi R_{\rm pipe})
\end{equation}
and find
\begin{equation}
\frac{Z_{\parallel}(k)}{Z_0} = \frac{jk}{2\pi R_{\rm pipe}} \int \! \!
\int_{\stackrel{\mbox{\scriptsize solid}}{\mbox{\scriptsize border}}} 
drdz \left( \frac{E_r - E_0}{E_0}\right).
\end{equation}

\section{Semi-Elliptical Interior Obstacle --- Iris}

We now proceed to calculate Eq.~(2.15) explicitly for the geometry of 
Fig.~1a. We change from the variable $r$ to the variable
\begin{equation}
y = R_{\rm pipe} - r,
\end{equation}
with $|y| \ll \lambda$, $|y| \ll R_{\rm pipe}$, and we use elliptical
coordinates \cite{7} defined by
\begin{equation}
\left. \begin{array}{l}y = c \cosh u \cos v, \ \ z = c \sinh u \sin v
\\ a = c \sinh u_0 \ , \ b = c \cosh u_0, \ \ c^2 = b^2 - a^2
\end{array}\right\}.
\end{equation}
The metric (Jacobian) is defined by
\begin{equation}
dzdy = c^2D(u,v) dudv,
\end{equation}
where
\begin{equation}
D(u,v) = \cosh^2 u \sin^2v + \sinh^2u \cos^2v,
\end{equation}
and the Laplacian operator can be written as
\begin{equation}
\frac{\partial^2}{\partial x^2} + \frac{\partial^2}{\partial y^2} =
\frac{1}{c^2 D(u,v)} \left( \frac{\partial^2}{\partial u^2} +
\frac{\partial^2}{\partial v^2}\right).
\end{equation}
The solution to Laplace's equation for the electrostatic potential in the
region $u \geq u_0$, with $\Psi (u_0, v) = 0$, and with the asymptotic
field
$E_0$, is
\begin{equation}
\Psi (u,v) = cE_0 \cos v [\cosh u - e^{u_0-u} \cosh u_0]
\end{equation}
where
\begin{eqnarray}
E_r &=& - \frac{\partial \Psi}{\partial r} = \frac{\partial \Psi}{\partial
y}
\nonumber \\
&=& E_0 - cE_0 e^{u_0} \cosh u_0 \frac{\partial}{\partial y} (\cos v
e^{-u}).
\end{eqnarray}
Here we choose $\cos v$ in the second term to preserve the symmetry around
$v=0$ ($z=0$), and to satisfy the boundary condition for all $v$ at $u=u_0$
and
for $v=\pm \pi /2$ with $u \geq u_0$.  Also we have
\begin{eqnarray}
\frac{\partial}{\partial y} &=& \frac{\partial u}{\partial y}
\frac{\partial}{\partial u} + \frac{\partial v}{\partial y}
\frac{\partial}{\partial v} \nonumber \\
&=& \frac{1}{cD(u,v)} \left[ \sinh u \cos v \frac{\partial}
{\partial u} - \cosh u \sin v \frac{\partial}{\partial v}\right].
\end{eqnarray}
Applying this to Eq.~(3.7) for $E_r - E_0$, we find from Eq.~(3.6) that
\begin{eqnarray}
&&\frac{Z_{\parallel}(k)}{Z_0} = \frac{-jkc^2}{2\pi R_{\rm pipe}} 
e^{u_0} \cosh u_0 \times \nonumber \\
&& \quad \int \! \! \int dudv \, [\cos^2v e^{-u} \sinh u - 
\sin^2 v e^{-u} \cosh u] \nonumber  \\  
&& = \frac{jkc^2e^{u_0}\cosh u_0}{2R_{\rm pipe}} \int \! \!
\int_{\stackrel{\mbox{\scriptsize solid}}{\mbox{\scriptsize border}}} 
 dudv \, [e^{-2u} - \cos 2v],
\end{eqnarray}
where we have used Eq.~(2.14).

We now let $Z \rightarrow \infty$ and cut off the integration over 
$v$ where
\begin{eqnarray}
y & = & c \cosh u_{\rm max} \cos v = Y, \quad \mbox{  or} 
 \nonumber \\
u_{\rm max} & = & \cosh^{-1} \frac{Y}{c \cos v} \gg 1 \ .
\end{eqnarray}
This leads to
\begin{eqnarray}
\frac{Z_{\parallel}(k)}{Z_0} &=& \frac{jkc^2e^{u_0} \cosh u_0}
{4\pi R_{\rm pipe}}
\int^{\frac{\pi}{2}}_{-\frac{\pi}{2}} dv
\int^{u_{\rm max}}_{u_0} du [e^{-2u} - \cos v] \nonumber \\
&\cong & \frac{jkc^2e^{u_0} \cosh u_0}{4\pi R_{\rm pipe}} \left[
\vphantom{\int^{\frac{\pi}{2}}_{-\frac{\pi}{2}}}
\frac{\pi}{2} e^{-2u_0}\right. \nonumber \\
&-& \left. \int^{\frac{\pi}{2}}_{-\frac{\pi}{2}} dv \cos 2v \left( \ln
\frac{2Y}{c \cos v} - u_0\right) \right],
\end{eqnarray}
or
\begin{equation}
\frac{Z_{\parallel}(k)}{Z_0} = \frac{jkc^2 \cosh^2 u_0}{4R_{\rm pipe}} =
\frac{jkb^2}{4R_{\rm pipe}}
\end{equation}
where the last integral in (3.11) was done by parts for $Y \gg c$.

For $a > b$ we need to modify our elliptical coordinates so that
\begin{equation}
\left. \begin{array}{l}y = c \sinh u \sin v, \ \ z = c \cosh u \cos v \\
a = c \cosh u_0 \ , \ b = c \sinh u_0, \ \ c^2 = a^2 - b^2
\end{array}\right\}.
\end{equation}
The matrix is unchanged, but now
\begin{equation}
\frac{\partial}{\partial y} = \frac{1}{D(u,v)} \left[ \frac{\cosh u 
\sin v}{c} \frac{\partial}{\partial u} + \frac{\sinh u \cos v}{c}
\frac{\partial}{\partial v}\right].
\end{equation}
This time one finds
\begin{equation}
\frac{Z_{\parallel}(k)}{Z_0} = \frac{jkc^2 \sinh^2 u_0}
{4R_{\rm pipe}} = \frac{jkb^2}{4R_{\rm pipe}},
\end{equation}
that is the result is unchanged from Eq.~(3.12), again depending only 
on the depth of the elliptical protrusion into the pipe and not on its 
width, as also found by Kurennoy \cite{1}.

\section{Semi-Elliptical Exterior Obstacle --- Cavity}

\subsection{Analytical Approach} 

We now turn to the exterior semi-elliptical obstacle in Fig.~1b.  
Here we need to choose appropriate potential forms for $u \leq u_0$, 
and for $u \geq u_0$, and match them at $u = u_0$.

We again start with Eq.~(2.15) and work with the complete set of 
solutions of the Laplace equation, namely $\cos nv \exp (\pm nu)$ and 
$\sin nv \exp (\pm nu)$. For $u \geq u_0$, (and $b > a$) with the 
asymptotic field $E_0$, we choose
\begin{equation}
\Psi (u,v) = E_0y - cE_0 \sum^{\infty}_{n=1} \alpha_n \cos nv e^{-nu}
\end{equation}
in order to satisfy the even symmetry about $v = 0$ ($z = 0$) and $\Psi
(u,\pm
\pi /2) = 0$ for $u \geq u_0$.  If we write
\begin{equation}
\Psi (u_0,v) = c E_0 f(v),
\end{equation}
where $f(v)$ is, as yet, an unknown function, we can solve for 
$\alpha_n$ in terms of $f(v)$ to obtain
\begin{equation}
\alpha_n = -\frac{2}{\pi} e^{nu_0} f_n + e^{u_0} \cosh u_0 \delta_{n1},
\end{equation}
where
\begin{equation}
f_n \equiv \int^{\frac{\pi}{2}}_{-\frac{\pi}{2}} dv \cos nv f(v).
\end{equation}

For $u \leq u_0$, we write
\begin{equation}
\Psi (u,v) = \sum^{\infty}_{m=1} \beta_n \cos nv \cosh mu
\end{equation}
for a potential which is well behaved within the ellipse.  
Recognizing in this case that $f(v) = 0$ for $\pi /2 < |v| < \pi$, 
we solve for $\beta_m$ in terms of $f(v)$ to obtain
\begin{equation}
\beta_m = \frac{f_m}{\pi \cosh mu_0},
\end{equation}
where $f_m$ is consistent with the definition in Eq.~(4.4).  Both odd 
and even values of $m$ must be included.

We now calculate the impedance as we did in the previous section, 
this time including the regions $u \geq u_0$, $|v| \leq \pi /2$ 
and $u \leq u_0$, $|v| \leq \pi$.  For  $u \geq u_0$ we find
\begin{eqnarray}
\frac{Z_{\parallel}^{(>)}(k)}{Z_0} \ = \ -\frac{jkc^2}
{4\pi R_{\rm pipe}} \int \! \! 
\int_{\stackrel{{\scriptsize u \geq u_0}}{{\scriptsize y < Y}}} dudv 
\sum^{\infty}_{\stackrel{{\scriptsize n=1}}{\mbox{\tiny odd}}} 
 n \alpha_n  \, \times \nonumber \\
\  \left[ e^{-(n-1) u} \cos (n+1) v - e^{-(n+1) u}
\cos (n-1) v\right].
\end{eqnarray}

Clearly, only the terms with $n=1$ survive, leading ultimately to
\begin{equation}
\frac{Z_{\parallel}^{(>)}(k)}{Z_0} = \frac{jkc^2}{4R_{\rm pipe}} 
 \alpha_1 e^{-u_0} \cosh u \ . 
\end{equation}

For $u \leq u_0$ we separate the two terms in Eq.~(2.15).  
The second is simply
\begin{equation}
\frac{Z_{\parallel ,2}^{(<)}(k)}{Z_0} = \frac{jkab}{\pi 4R_{\rm pipe}} 
= \frac{jkc^2 \cosh u_0 \sinh u_0}{4R_{\rm pipe}}
\end{equation}
The first is
\begin{eqnarray}
\frac{Z_{\parallel ,1}^{(<)}(k)}{Z_0} 
 = -\frac{jkc^2}{4\pi R_{\rm pipe}}
 \int \! \! \int_{u \leq u_0} dudv \nonumber \\
\ \times \sum^{\infty}_{m=1} m\beta_m [\cosh (m+1) u \cos
(m-1) v \nonumber \\
\qquad + \cosh (m-1) u \cos (m+1) v].
\end{eqnarray}
Again, only the term $m=1$ survives, and is
\begin{equation}
\frac{Z_{\parallel ,1}^{(<)}(k)}{Z_0} = -\frac{jkc^2}{4R_{\rm pipe}}
\beta_1
\cosh u_0 \sinh u_0.
\end{equation}
Using Eqs.~(4.3) and (4.6) we have for the impedance
\begin{equation}
\frac{Z_{\parallel}(k)}{Z_0} = \frac{jk}{2R_{\rm pipe}} 
\left[ \frac{b^2}{2} + ab - 
 (a+b)^2 \frac{f_1}{\pi} e^{-u_0}\right].
\end{equation}

In order to find $f_1$, we must obtain and solve the integral 
equation which represents the match of $\partial \Psi /\partial u$ 
at $u = u_0$, $|v| \leq \pi /2$.  Here
\begin{eqnarray}
\left. \frac{\partial \Psi}{\partial u}\right|_{u = u_{o^+}} 
&=& cE_0 \left[
\vphantom{\sum^{\infty}_{\stackrel{n=1}{\rm odd}}} \sinh u_0 \cos v 
\right. \nonumber \\ 
&+& \left. \sum^{\infty}_{\stackrel{{\scriptsize n=1}}
{\mbox {\tiny odd}}} n \alpha_n e^{-nu_0} \cosh v\right]
\end{eqnarray}
and
\begin{equation}
\left. \frac{\partial \Psi}{\partial u}\right|_{u = u_{o^-}} = cE_0
\sum^{\infty}_{m=1} m \beta_m \sinh mu_0 \cos mv.
\end{equation}
Equating Eqs.~(4.13) and (4.14), and using Eqs.~(4.3) and (4.6), we find
\begin{equation}
\int^{\frac{\pi}{2}}_{-\frac{\pi}{2}} dv^{\prime} f(v^{\prime})
K(v,v^{\prime}) = \pi \cos v e^{u_0}
\end{equation}
where
\begin{eqnarray}
K(v,v^{\prime}) &=& \sum^{\infty}_{\stackrel{{\scriptsize n=1}}
{\mbox{\tiny odd}}} (2 + \tanh nu_0) \, n
\cos nv \cos uv^{\prime} \nonumber \\ 
&+& \sum^{\infty}_{\stackrel{{\scriptsize m=2}}
{\mbox{\tiny even}}} m\tanh mu_0 \cos mv \cos mv.
\end{eqnarray}
We now multiply Eq.~(4.15) by 
$\int^{\frac{\pi}{2}}_{-\frac{\pi}{2}} dv f(v)$ to obtain
\begin{eqnarray}
\frac{\left[ \int^{\frac{\pi}{2}}_{-\frac{\pi}{2}} dvf(v) \cos
v\right]^2}{\int^{\frac{\pi}{2}}_{-\frac{\pi}{2}} dvf(v)
\int^{\frac{\pi}{2}}_{-\frac{\pi}{2}} dv^{\prime} f(v^{\prime})
K(v,v^{\prime})} &=& \frac{\int^{\frac{\pi}{2}}_{-\frac{\pi}{2}} 
dvf(v)\cos v}{\pi e^{u_0}} \nonumber \\
&=& \frac{f_1e^{-u_0}}{\pi}.
\end{eqnarray}
This is a variational form for $f_1$, the only unknown parameter 
in Eq.~(4.12) for the impedance.  An accurate numerical value for 
$Z_{\parallel} (k)/Z_0$ can be found by expanding $f(v)$ into a 
complete set in the interval $|v| \leq \pi /2$, then truncating and 
solving the resulting matrix equations obtained by
maximizing Eq.~(4.17).  We write
\begin{equation}
f(v) = \sum^P_{\stackrel{{\scriptsize p=1}}
{\mbox{\tiny odd}}} c_p \sin \left(
p\frac{\pi}{2}\right) \frac{\cos pv}{p},
\end{equation}
truncated at $p=P$ and normalize $f(v)$ so that $c_1=1$. This leads to
\begin{eqnarray}
\frac{f_1e^{-u_o}}{\pi} = \left[ \vphantom{\sum^{P}_{\stackrel{p=3}
{\rm odd}}} H_{11} \right.
&+&  2\sum^{P}_{\stackrel{{\scriptsize p=3}}{\mbox{\tiny odd}}} 
c_p H_{p1} \nonumber \\   &+& \left. 
\sum^{P}_{\stackrel{{\scriptsize p=3}}{\mbox{\tiny odd}}} 
\sum^{P}_{\stackrel{{\scriptsize q=3}}{\mbox{\tiny odd}}}
c_pc_qH_{pq}\right]^{-1},
\end{eqnarray}
where the symmetric matrix $H_{pq}$ is
\begin{eqnarray}
H_{pq} = H_{qp} &=& \frac{2+\tanh pu_0}{P} \delta_{pq} \nonumber \\
&+& \frac{16}{\pi^2} \sum^{\infty}_{\stackrel{m=2}{\mbox{\tiny even}}} 
\frac{m\tanh mu_0}{(m^2-p^2)(m^2-q^2)}.
\end{eqnarray}
Maximizing Eq.~(4.19) with respect to the coefficients $c_p$, 
$p=3,5,\cdots,P$, leads to
\begin{equation}
\frac{f_1e^{-u_o}}{\pi} = \left[ H_{11} -
\sum^{P}_{\stackrel{{\scriptsize p=3}}{\mbox{\tiny odd}}} 
\sum^{P}_{\stackrel{{\scriptsize q=3}}{\mbox{\tiny odd}}}
H_{1p} (H^{-1})_{pq} H_{q1}\right]^{-1}.
\end{equation}
Here $(H^{-1})_{pq}$ is the inverse of the matrix $H_{pq}$ with $p$ 
and $q=3,5,7\cdots P$.  This square matrix has the dimension 
\begin{equation}
(P-1)/2 \ {\rm by} \ (P-1)/2.
\end{equation}
Note that
\begin{equation}
\tanh pu_0 = \left(1-w^p \right) / \left(1+w^p \right) 
\end{equation}
and
\begin{equation}
\tanh mu_0 = \left(1-w^m \right) / \left(1+w^m \right) 
\end{equation}
with
\begin{equation}
w =  (b - a) / (b + a) \ .
\end{equation}
The final  result for the impedance is given in Eq.~(4.12), using 
Eqs.~(4.20) and (4.21).

The analysis for an obstacle with $a >b$ proceeds in a similar, but 
not identical pattern.  The result is once again Eq.~(4.12), using 
Eqs.~(4.20) and (4.21), with only one change in Eq.~(4.20):
\[
\tanh pu_0 \rightarrow \coth pu_0,
\]
with $\tanh u_0$ now being $a/b$ instead of $b/a$.  Thus $w$ in 
Eq.~(4.25) is replaced by $-w$ in Eq.~(4.23), but the use of 
$\coth pu_0$ instead of $\tanh pu_0$ for odd $p$ leaves the 
expression for $H_{pq}$ in terms of $a$ and $b$ unchanged.  
The same is true for the term $\tanh mu_0$ in Eq.~(4.24) since
$m$ is even.  So the final expression in Eq.~(4.12) is unchanged 
provided $H_{pq}$ in
Eqs.~(4.20) and (4.21) is expressed in terms of $a$ and $b$.

\subsection{Variational Approach - Numerical Results}

We proceed with a numerical investigation of the variational 
scheme described by Eqs.~(4.12)-(4.21). Truncating the sum in  
the denominator of Eq.~(4.21) at different $N=(P-1)/2=1,2,3,\ldots$, 
we explore the scheme  convergence, and compare the results for the 
impedance (4.12) with  those obtained by other methods. In doing so, 
it is convenient to rewrite Eq.~(4.12) in the following form 
\begin{equation}
\frac{Z_\|(k)}{Z_0} = \frac{jk}{2 \pi R_{\rm pipe}} \frac{\pi ab}{2}
F\left(\frac{a}{b} \right), 
\end{equation}
where
\begin{equation}
F(x) = \frac{1}{x} + 2 - \frac{2(1/x+2+x)}{2+x+16 s(x)/\pi^2- \Sigma(x)},
\end{equation}
and 
\begin{equation}
s(x) = \frac{1}{8}\sum_{n=1}^\infty \frac{n}{(n^2-1/4)^2}
\frac{(1+x)^{2n}-(1-x)^{2n}}{(1+x)^{2n}+(1-x)^{2n}}. 
\end{equation}
Here $\Sigma(x)$ denotes the sum in the denominator of Eq.~(4.21) 
which is to be truncated.

The advantage of the representation (4.26) is that we know the 
asymptotic behavior of $F(x)$ for two limiting cases.  
For $x \ll 1$, i.e., when $a \ll b$, but still $b \ll R_{\rm pipe}$  
--- a short and deep enlargement --- it has been demonstrated 
in \cite{8} that 
\begin{equation}
F(x) \to 1 - \frac{4}{\pi^2}x.   
\end{equation}
In this limit, the inductive impedance in Eq.~(4.26) is mostly of 
magnetic origin: the beam magnetic field fills the cavity  volume 
without being substantially perturbed, and therefore  the inductance 
is simply proportional to the area of the obstacle cross section. 
A correction of the order of $x=a/b$ to this term comes from the 
electric contribution. For a deep pillbox of depth $h$
which is much larger than width $g$, the electric contribution was
calculated in \cite{8} by means of conformal mapping. It results 
in the electric term $-g^2/(2\pi)$, which is small compared to the 
magnetic one, equal to $hg$ for such a pillbox. 
Obviously, the  shape of a short and deep enlargement ---
rectangular or  semi-elliptical --- does not affect the electric 
term as long as $g \ll h$, since the beam electric field does not 
penetrate deeply into such a cavity, unlike the magnetic one. 
Substituting $g=2a$ into the electric term, and replacing the 
pillbox area $hg$ by the semi-ellipse area $\pi ab/2$ leads to
the asymptotic form in Eq.~(4.29). 

The opposite limit, $x \gg 1$, corresponds to a very shallow cavity, 
$a \ll b$. It has been shown for many particular  shapes of such 
cavities (see \cite{8} and references therein) that the 
low-frequency impedance of a small shallow cavity of the depth $h$ 
and of an iris with the same cross section and having the
same depth, are both inductive, equal to each other, and in the 
leading order are proportional to $h^2$.  Since we already know the 
answer for a semi-elliptical iris (see Section III) we expect that 
for $x \gg 1$ 
\begin{equation}
 F(x) \to \frac{1}{x},
\end{equation}
to match the low-frequency impedance of the shallow iris, given in 
Eq.~(3.15).

\begin{figure}[hbt]
\centerline{\psfig{file=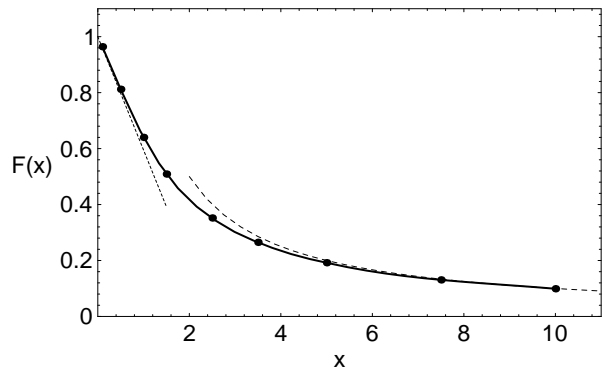,width=8.5cm}}
\caption{Function $F(x)$ versus ellipse aspect ratio $x=a/b$.  
The solid curve is an interpolation of numerical results (thick 
points).  The dashed lines show the asymptotic behavior Eqs.\ 
(4.29) and (4.30).}
\end{figure}

The results of our numerical study are shown in Fig.~2, where the 
function $F(x)$ is plotted against the ellipse aspect ratio $x=a/b$. 
The convergence of the variational scheme is rather fast for all 
values of $x$; in fact, results obtained with $N=1$ (i.e., when the 
matrix in the sum $\Sigma(x)$ is truncated to merely a single number) 
and those for $N=8$, when the matrix has size $8\times 8$, differ by 
less than 0.5\%. And, of  course, we can obtain the asymptotic
value of $F(x)$ for large $N$ with much better accuracy, well below
$10^{-3}$, simply by  extrapolating the results for different matrix 
sizes at fixed $x$. Figure~2 also shows very good agreement with the 
expected asymptotic behavior Eq.~(4.29) for small $x$ and Eq.~(4.30) 
for large $x$.

\section{Transverse Coupling Impedance}

We start with a dipole drive current for the transverse impedance 
in the form \cite{9}
\begin{eqnarray}
J_z(x,y,z;k) = &&I_0 \delta (y) \exp (-jkz) \times \nonumber \\
&&[\delta (x-x_1) - \delta (x + x_1)],
\end{eqnarray}
where we eventually proceed to the limit $x_1 \rightarrow 0$.  
It is straightforward to show \cite{9} that the transverse 
impedance in the $x$ direction can be written as
\begin{equation}
Z_x(k) = -\frac{1}{4kx^2_1 |I_0|^2} \int dv \mbox{\boldmath $E$} 
\cdot \mbox{\boldmath $J$}^*,
\end{equation}
analogous to Eq.~(2.1) for the longitudinal impedance.  Use of 
Maxwell's equations as we did in Section II, leads to
\begin{equation}
Z_x(k) = -\frac{1}{4kx^2_1 |I_0|^2} \int_{S_2 \neq S_1} dS_z
\mbox{\boldmath
$n$}_2 \cdot \mbox{\boldmath $E$}_1^* \times \mbox{\boldmath $H$}_2,
\end{equation}
but we must now use the form of $\mbox{\boldmath $E$}_1$ 
(and $\mbox{\boldmath $H$}_2$) appropriate to the source current in 
Eq.~(4.26).  In fact, we now have for $\mbox{\boldmath $E$}_1$ and 
$Z_0\mbox{\boldmath $H$}_1$ at the beam pipe wall
\begin{equation}
E_{1r} = Z_0 H_{1\phi} = \frac{2Z_0I_0}{\pi R^2_{\rm pipe}} x_1 \cos 
\phi \exp(-jkz)
\end{equation}
\[
E_{1\phi} = Z_0 H_{1r} = 0.
\]
As a result, we can write 
\begin{eqnarray}
&&\frac{Z_x(k)}{Z_0} \cong -\frac{1}{2kx_1I_0Z_0 \pi R_{\rm pipe}} 
 \times \nonumber \\
&&\int d\phi \cos \phi \int dr (Z_0 H_{2\phi} e^{jkz}).
\end{eqnarray}
Once again we have written the impedance as an integral along the 
surface of the obstacle, where $H_{2\phi}$ arises from the driving 
field components $E_{1r}$ and $H_{1\phi}$ at the wall.  Dropping 
the subscript 2, extracting the factor $\cos \phi$ from $H_{\phi}$ 
and integrating over $\phi$ leads to
\begin{equation}
\frac{Z_x(k)}{Z_0} \cong -\frac{1}{2kx_1I_0Z_0R_{\rm pipe}} \int dr
(Z_0H_{\phi}
e^{jkz}).
\end{equation}

We now write Eq.~(5.6) as a double integral over $drdz$ as we did 
in Section II, obtaining
\begin{equation}
\frac{Z_x(k)}{Z_0} = \frac{-j}{R^3_{\rm pipe}} 
\int_{\stackrel{\mbox{\scriptsize solid}}{\mbox{\scriptsize border}}} 
dr \int dz \left( \frac{E_r - E_0}{E_0}\right) 
\end{equation}
where $E_0$, the maximum asymptotic field at the wall, is
\begin{equation}
E_0 = \frac{2Z_0I_0}{\pi R^2_{\rm pipe}} x_1.
\end{equation}

Comparison of Eq.~(5.8 ) with Eq.~(2.15) shows that the calculations 
for an exterior and an interior obstacle are exactly the same as 
they were for the longitudinal impedance.  In fact, the results for 
the transverse impedance can be obtained simply by multiplying the 
results for the longitudinal impedance in
Eqs.~(3.12), (3.15), (4.12) and (4.21) by $2/kR^2_{\rm pipe}$.

\end{document}